\documentclass{emulateapj}
\usepackage{apjfonts}

\begin{document} 
\title{Imaging of the Radio Remnant of SN 1987A at 12 mm wavelength}
\author{R. N. Manchester\altaffilmark{1},
B. M. Gaensler\altaffilmark{2}, L. Staveley-Smith\altaffilmark{1}, 
M. J. Kesteven\altaffilmark{1} and A. K. Tzioumis\altaffilmark{1}}
\altaffiltext{1}{Australia Telescope National Facility, CSIRO, PO Box
76, Epping NSW 1710, Australia} \altaffiltext{2}{Harvard-Smithsonian
Center for Astrophysics, 60 Garden Street, MS-6, Cambridge, MA 02138}
\email{dick.manchester@csiro.au}

\begin{abstract} 
Observations of the radio remnant of Supernova 1987A using the
Australia Telescope Compact Array in the 12-mm band on 2003 July 31
(day 6002.7 after the explosion) give the first fully resolved radio
image of the supernova remnant. The diffraction-limited image has a
resolution of about 0\farcs45, a factor of two better than that of the
3-cm images previously obtained. There is excellent agreement between
the 12-mm image and a contemporaneous super-resolved 3-cm
image. Super-resolution of the 12-mm image gives a further factor of
two improvement in resolution, to $0\farcs25$, albeit with limited
dynamic range.  While the spatial distributions of the radio and X-ray
emission are broadly similar, there are significant differences in
detail with no correspondence in the regions of brightest
emission. The 12-mm image is well modelled by a thick equatorial ring
inclined at $43\degr$ to the line of sight. This, together with the
common east-west asymmetry and the relatively steady increase in the
radio flux density, suggests that the reverse shock is the main
site for generation of the radio emission.
\end{abstract}


\section{Introduction}
Supernova 1987A, the brightest and closest supernova (SN) in nearly 400
years, was first observed on 1987 February 24. The supernova, located
in the Large Magellanic Cloud, illuminated and ionized much of the
circumstellar material, resulting in the dramatic optical rings seen
in recombination-line emission \citep{bkh+95,slc+02} and scattered light
\citep[e.g.,][]{ckh95,sck+05}. {\it Hubble Space Telescope} imaging has revealed a
series of ``hot spots'' around the inner edge of the central
equatorial ring, indicating that the SN blast wave is
beginning to impact on the denser circumstellar material of the
equatorial ring \citep{slc+02}. Spectral imaging of high-velocity
($\sim 12,000$ km s$^{-1}$) Ly$\alpha$ and H$\alpha$ emission has enabled
mapping of the reverse shock formed behind the blast wave
\citep{mmc+03}. These observations show that in 2000 -- 2001, around
5000 days after the SN, the reverse shock had a radius of about 75\%
that of the equatorial ring and was expanding at about 80\% of the
blast-wave velocity. The shock appears about 40\% brighter in the
eastern quadrant and is concentrated within $30\degr$ of the equatorial
plane.

SN 1987A was observed as a short-lived radio supernova in the weeks
following the optical supernova \citep{tcb+87}. It was redetected in
mid-1990 (around day 1200) when increasing radio emission signalled
the birth of the supernova remnant (SNR) \citep{smk+92}. Observations
with the Molonglo Observatory Synthesis Telescope and Australia
Telescope Compact Array (ATCA) show that, at least up until the
beginning of 2001, the SNR radio emission has increased nearly
linearly with time at frequencies between 843 MHz and 9 GHz
\citep{bch+01,mgw+02}. The radio emission appears to be optically thin
synchrotron emission with a power-law spectrum of index about
$-0.9$. From late-1992 the SNR has been strong enough to resolve at 9
GHz, showing that radio emission is annular in form, contained within
the optical equatorial ring and expanding at about 3000 km s$^{-1}$
\citep{sbr+93,gms+97,mgw+02}. The radio emission peaks on the eastern and
western sides of the ring and is signficantly brighter on the eastern
side. These images had an effective resolution of about
0\farcs5 and relied on the use of ``super-resolution'' techniques in
which visibilities in the outer parts of the sampled $u-v$ plane are
given higher weight. The resulting images are inherently uncertain
since the solution extrapolates into unsampled parts of the uv-plane.

X-ray emission was first detected from SN 1987A about 1500 days after
the explosion \citep{ght94,bbp94}. Recent observations with the {\it
Chandra X-ray Observatory} \citep{pzb+04,pzb+05} show that the X-ray
luminosity is increasing dramatically with a doubling time of about
600 days. {\it Chandra} imaging observations taken in 2000 December
\citep{pbg+02} showed different spatial distributions for the lower
energy (0.3 -- 1.2 keV) photons and those at higher energies (1.2 -- 8
keV). The lower energy emission was concentrated in hot spots around
the inner edge of the equatorial ring and was closely correlated with the
optical emission, whereas the higher energy emission was slightly more
central and apparently related to the radio emission with a pronounced
east-west asymmetry. In the more recent images, this distinction
appears to have disappeared with the emission more uniformly
distributed around the ring at all energies.

Receivers operating in the 12-mm band (16 -- 26 GHz) were installed on
three antennas of the ATCA in 2001 October and on the remaining three
antennas in 2003 April, giving a 6-km maximum baseline. One of the
major scientific justifications for development of the 12-mm system
was to obtain radio images of SNR 1987A with a resolution comparable
to those obtained by the {\it Hubble Space Telescope} and the {\it
Chandra X-ray Observatory}. SNR 1987A was observed at 12~mm on 2001
October 27 (day 5360 after the SN) with a 1.5-km maximum
baseline. While these observations just marginally resolved the SNR,
they gave flux densities of $20\pm 1$ and $18\pm 1$ mJy at 17 and 19
GHz, respectively \citep{mg01}. In this paper we report on 12-mm
observations of SNR 1987A made on 2003 July 31 (day 6002.7) with six
antennas over the full 6-km baseline in exceptionally stable
atmospheric conditions.

\section{Observations and Results}

SNR 1987A was observed for 11 hours on 2003 July 31 using the 6D
configuration of the ATCA with 12-mm receivers on each of the six
antennas. Observations were obtained in two bands simultaneously, each
of 128 MHz bandwidth, and centered on frequencies of 17.34 and 19.65
GHz respectively. Two phase calibrators were observed: 0541-7332 for 2
min at approximately 10-min intervals and 0516-621 for 2 min at
approximately hourly intervals. Visibility phases for the secondary
calibrators were relatively stable on all baselines with maximum drift
rates on 6-km baselines of about $50\degr$ per hour.

All data processing was done using the
\anchor{http://www.atnf.csiro.au/computing/software/miriad}{{\sc
miriad}}\footnote{See
http://www.atnf.csiro.au/computing/software/miriad} synthesis data
analysis system. Positions were referred to the \citet{mae+98} J2000
position for 0516$-$621: $05^{\rm h} 16^{\rm m} 44\fs9262(5)$,
$-62\degr 07' 05\farcs398(5)$ (where the uncertainty in the last
quoted digit is given in parentheses) derived from very long baseline
interferometry (VLBI) observations. The derived J2000 position for
0541$-$7332 was $05^{\rm h} 41^{\rm m} 50\fs7669(8)$, $-73\degr 32'
15\farcs340(8)$. Flux densities were based on observations of Mars
which was assumed to have a disk temperature of 190 K, giving a flux
density of 16.0 Jy at 17.34 GHz and 20.5 Jy at 19.65 GHz on the date
of the observations.  Mars was resolved at all except very short
baselines. The flux density scale was transferred to the primary
calibrator 0826$-$373 using the {\sc miriad} task {\sc plboot} giving
flux densities for this source of 1.75 Jy and 1.60 Jy at 17.34 and
19.65 GHz respectively.

\begin{figure*}
\plottwo{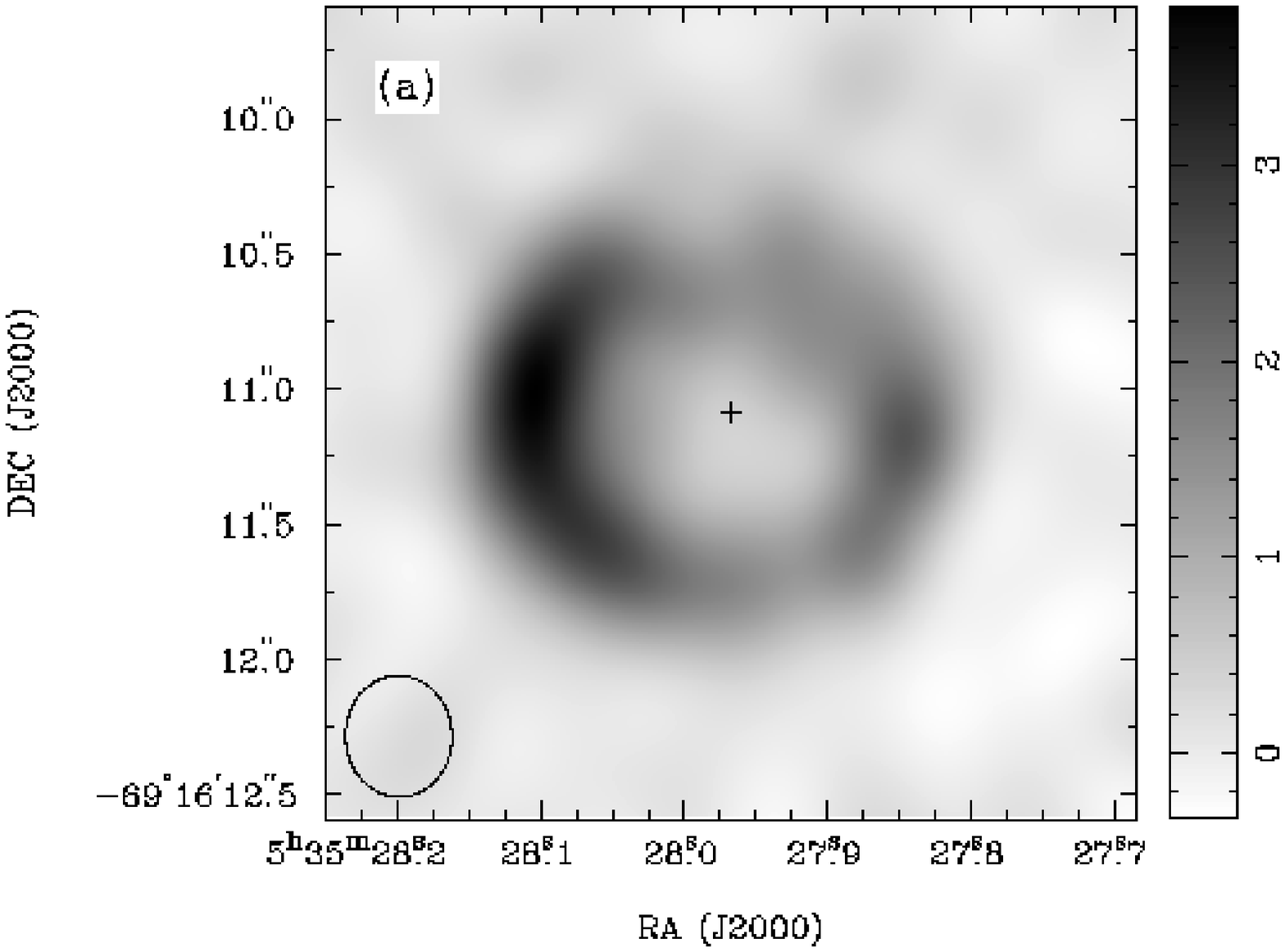}{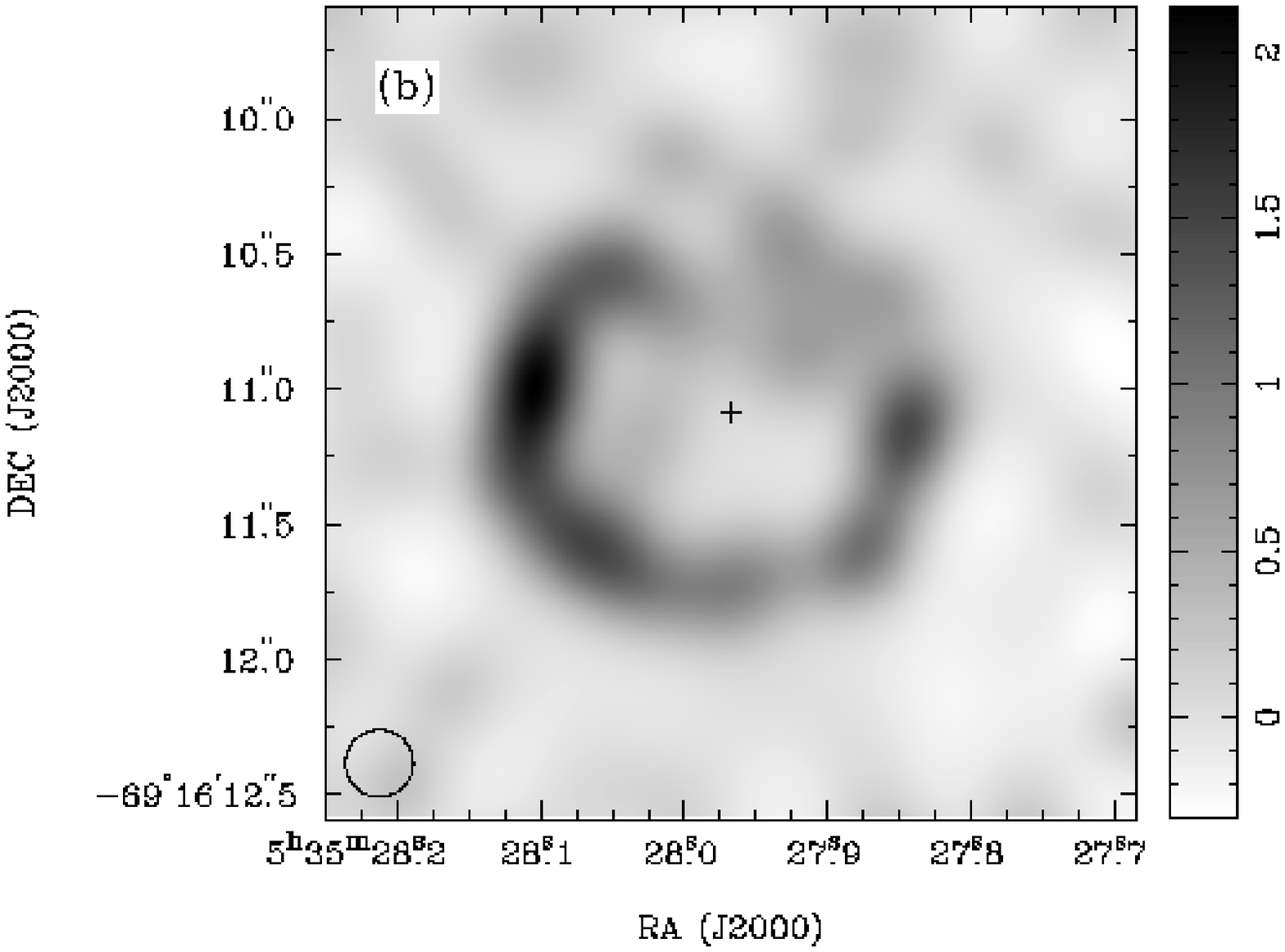}
\caption{Diffraction-limited (a) and super-resolved (b) images
  of SNR 1987A at 12~mm wavelength from observations made on 2003 July
  31 using the ATCA. The central cross marks the VLBI position of
  SN~1987A and the flux density scale is in units of mJy/beam. For the
  diffraction-limited image, the off-source rms noise is approximately
  120 $\mu$Jy. The half-maximum contour of the restoring beam is shown
  in the lower-left corner of each plot.}
\label{fg:12mm}
\end{figure*}

Figure~\ref{fg:12mm}a shows the 12-mm image obtained by inverting the
calibrated $u-v$ data for both observed frequencies with uniform
weighting. The resulting image was cleaned and restored with the
diffraction-limited beam of half-maximum size $0\farcs45 \times
0\farcs39$ with the major axis at position angle 2\degr (measured from
north through east). Positions are absolute and the cross marks the
position for SN 1987A determined by \citet{rjs+95} by relating the
VLBI and Hipparchos reference frames; this position has an uncertainty
of about 0\farcs07.  A super-resolved version of this image, obtained
using the {\sc maxen} routine and restored with a 0\farcs25 circular
beam is shown in Figure~\ref{fg:12mm}b. Because of the relatively low
signal-to-noise ratio of the 12-mm data, the dynamic range of this
super-resolved image is only about 8:1.

These observations give an integrated flux density for SNR 1987A of
28.5 mJy and 24.5 mJy at 17.3 GHz and 19.6 GHz respectively, with an
estimated uncertainty of 1.5 mJy in each case. A 12-hour ATCA observation at
8.6 GHz on 2003 August 1 gave an integrated flux density for the SNR
of $46 \pm 1$ mJy. Taken at face value, these results imply a
mean spectral index at day 6003 of $-0.7$ with an uncertainty of about
0.1. 

\section{Discussion}
\begin{figure}
\plotone{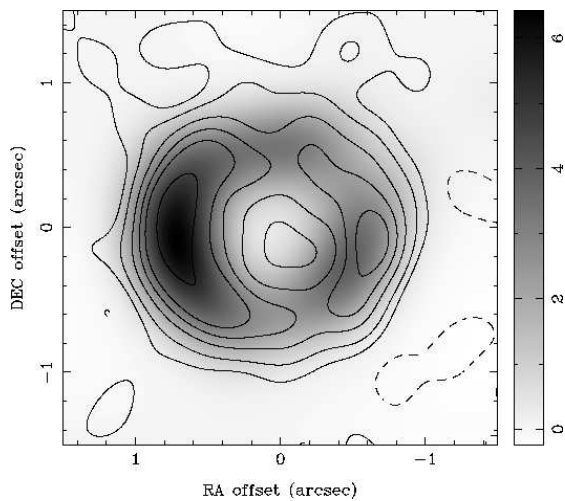}
\caption{Overlay of the diffraction-limited 12mm image (contours) on
  a super-resolved 3cm image from ATCA observations on 2003 August 1
  (greyscale). The contour intervals are -0.2, 0.2, 0.5, 1, 1.5, 2 and
  3 mJy/beam and the greyscale wedge is labelled in units of
  mJy/beam.}
\label{fg:3cm_12mm}
\end{figure}

Figure~\ref{fg:12mm}a shows that the SNR ring structure is clearly
resolved with the diffraction-limited 12-mm imaging. Previously, with
the 3-cm imaging \citep{gms+97,mgw+02}, super-resolution techniques
were required to reveal this structure. While the consistency of
successive super-resolved images of SNR 1987A gave us confidence that
they were reliable, a direct confirmation of that is valuable. This is
provided by Figure~\ref{fg:3cm_12mm} which shows an overlay of the
12-mm diffraction-limited image on the 3-cm super-resolved image based
on the ATCA observations of 2003 August 1. Although there is an
excellent overall correlation between the two images, there are slight
differences. On both the eastern and western sides, the peak 3-cm
intensities are displaced slightly southward of the 12-mm
peaks. Whether or not this indicates different spectral indices in
different parts of the ring remains to be confirmed by future
observations.

\begin{figure}
\plotone{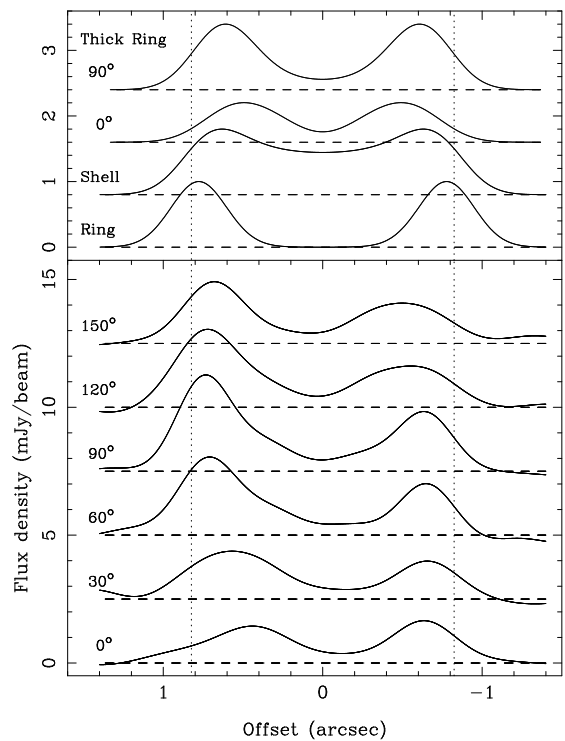}
\caption{Slices through the 12-mm diffraction-limited image at six
position angles. Dashed lines show the zero for each slice. The offset
is radial from the VLBI position and is positive toward north and/or
east. The vertical dotted lines show the radius of a thin shell fitted
to the $u-v$ data. Slices through a thin ring, a thin shell and an
inclined thick ring are shown in the upper part of the figure -- see
text for details.  }
\label{fg:slice}
\end{figure}

Figure~\ref{fg:slice} shows slices through the diffraction-limited
12-mm image (Figure~\ref{fg:12mm}a) at six position angles. These
clearly show the pronounced east-west asymmetry with the eastern lobe
stronger and centered at a slightly greater radial distance compared
to the western lobe. It is also clear that the north-south diameter of
the radio emission is less than the east-west diameter; based on the
separation of the slice maxima, the ratio of diameters is about
0.75. Within the uncertainties, this ellipticity is consistent with
the inclination of $43\degr$ derived for the optical equatorial ring
\citep{ckh95}.

Emission at the centre of the ring is significant and about 20\% of
the average ring intensity. Images were generated within {\sc miriad}
of two disks and two optically thin spheres of diameter $1\farcs55$
and $1\farcs65$ respectively and differenced to form a ring and shell
of thickness $0\farcs1$ in each case. These were then convolved with
the diffraction-limited beam. Slices through these model images are
shown in the upper part of Figure~\ref{fg:slice}. Clearly, even a thin
shell of emission has a larger surface brightness at its center than
the observed SNR, whereas the thin ring or annulus has a lower (zero)
surface brightness at its center. This suggests that something
between a shell and an equatorial ring is the best model. The reverse
shock described by \citet{mmc+03} has just this morphology. The top
two curves in Figure~\ref{fg:slice} are slices at position angles
$0\degr$ and $90\degr$ through an optically-thin equatorial ring of
latitude extent $\pm 30\degr$ and inner and outer radii of $0\farcs6$
and $0\farcs8$ respectively, inclined at $43\degr$ to the line of
sight. Apart from the unmodelled east-west asymmetry, these model
slices agree remarkably well with the corresponding cuts through the
12-mm image (lower part of Figure~\ref{fg:slice}). 

In previous analyses, the size and expansion rate of the radio shell
has been measured by fitting an optically thin shell to the 3-cm $u-v$
data \citep{gms+97,mgw+02}.  The diameter obtained from such a fit to
the 12-mm data is $1\farcs649 \pm 0\farcs004$, shown by the vertical
dotted lines in Figure~\ref{fg:slice}. The model slice for a thin
shell shows that the emission maxima lie significantly inside the
nominal shell radius, consistent with the observed profiles. A similar
fit to the 3-cm data of 2003 August 1 gives a shell diameter of
$1\farcs652\pm 0\farcs003$, fully consistent with the fit to the 12-mm
data. 

\begin{figure}
\plotone{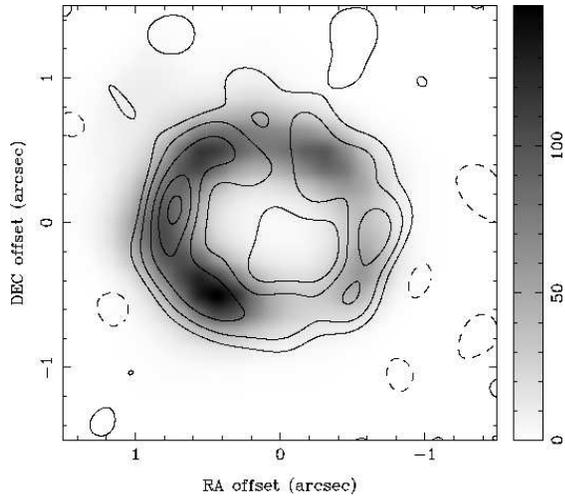}
\caption{Overlay of the super-resolved 12mm image (contours) on a {\it
  Chandra} 0.3 -- 8 keV image taken on 2003 July 8 (greyscale). The
  contour intervals are -0.2, 0.2, 0.5, 1, 1.5 and 2 mJy/beam and the
  greyscale is linear in X-ray counts after deconvolution and
  smoothing.}
\label{fg:Chandra_12mm}
\end{figure}

Recent X-ray observations \citep{pzb+04,pzb+05} show new hot spots
emerging on the western side of the ring at both high and low
energies, with the X-ray emission now forming a more complete
ring. Figure~\ref{fg:Chandra_12mm} shows an overlay of the
super-resolved 12-mm image aligned by eye with a contemporaneous
broad-band X-ray image \citep{pzb+05}. While the overall annular
distribution of emission is very similar in the two images, in detail
they differ substantially. The radio emission peaks on the eastern and
western sides of the ring whereas the X-ray emission tends to peak in
the ``corners'' of the ring with saddle points at the radio
peaks. {\it Hubble Space Telescope} images of SNR 1987A taken at a
similar epoch in H$\alpha$ light \citep[reproduced in the review
by][]{mcc05} show a large number of hot spots distributed around the
inner edge of the equatorial ring. Again, there is no detailed
correspondence with either the radio or X-ray emission distribution.

It is clear that the physical conditions under which the radio
emission, optical (H$\alpha$) hot spots and the X-ray emission are
generated are different, although all result from the interaction of
the expanding supernova ejecta with circumstellar material. The
H$\alpha$ hot spots are clearly associated with the inner edge of the
equatorial ring and have little or no azimuthal asymmetry. The radio
emission is concentrated on the eastern and western sides and has a
pronounced east-west asymmetry, with a much brighter eastern
lobe. Initially the X-ray emission shared some of these features but,
with the recent rapid brightening, it more resembles the optical
distribution. The time evolution of the intensity of the optical and
X-ray emission is also similar, being very non-linear with a rapidly
increasing gradient, whereas the radio emission has increased almost
linearly over the past decade with just a very small increase in
gradient \citep{mgw+02}. 

As discussed by previous authors \citep[e.g.,][and references
therein]{slc+02,pzb+04,mcc05}, these results suggest that the optical
emission and at least the rapidly increasing X-ray emission are
generated by the interaction of the supernova blast wave with
dense regions of the equatorial ring which extend inward. The fact
that the emission is now increasing rapidly and becoming more uniform
around the ring suggests that the blast wave is now encountering the
main ring structure. Furthermore, the similar spatial distribution and
similarly short evolution timescales shows that the optical and X-ray
emitting regions coexist behind the advancing shock front and are
generated contemporaneously.

In contrast, the radio emission began to increase at an earlier time
and has increased more-or-less steadily since. This suggests that the
relativistic electrons responsible for the synchrotron emission are
more widely distributed behind the shock front, perhaps filling the
region between the forward and reverse shocks. It is possible that the
reverse shock is the main site for acceleration of the
synchrotron-emitting electrons. This idea is supported by the
east-west asymmetry common to the radio emission and the reverse shock
and by excellent agreement of the model slices in
Figure~\ref{fg:slice} based on this morphology with the corresponding
slices through the observed image. The lifetime of the sychrotron
electrons is very long. Hence, unless adiabatic expansion or other
loss processes are important, the rate of generation of
synchrotron-emitting electrons has been relatively stable since day
1200 when the radio emission was first detected. At least at recent
times, the area of the shock surface (whether it is the forward or the
reverse shock) has been growing roughly as $t^2$, suggesting that the
rate of particle acceleration per unit area of the shock front has
been declining at roughly this rate.

The measured flux densities of the SNR at 17 and 19 GHz are above the
values extrapolated from lower frequencies. The SNR spectrum measured
between 0.8 and 10 GHz is becoming flatter, with a gradient of roughly
$4\times 10^{-5}$ per day \citep{mgw+02}. The observed slope at day
6000 is about $-0.8$ \citep{smg+05}. This is marginally consistent
with the value derived from the measured 12-mm and 3-cm flux densities
at this time, $-0.7 \pm 0.1$, suggesting that the rate of spectral
flattening may be increasing or that the radio spectrum may not be
power-law over the whole range from 0.8 to 19 GHz. We note that
diffusive shock acceleration theories predict a concave synchrotron
spectrum and there is some evidence for this in other SNRs
\citep[e.g.,][]{re92,jrdb03}.

The increasing radio flux density of SNR 1987A will soon give
super-resolved images with higher resolution and improved dynamic
range, allowing a more detailed comparison with the optical and X-ray
images. This should elucidate the relationship between the emission
processes in the various bands, showing whether or not the
acceleration of the radio sychrotron electrons occurs at the reverse
shock as suggested above. The increasing flux density will also allow
a more precise measurement of the frequency and spatial dependence of
the radio spectral index.

\section*{Acknowledgments}
We thank Dr Sangwook Park for supplying the {\it Chandra} 0.3 -- 8 keV
X-ray image of SN 1987A. The Australia Telescope Compact Array is part
of the Australia Telescope which is funded by the Commonwealth of
Australia for operation as a National Facility managed by CSIRO. We
especially thank the ATNF engineers and other staff responsible for
the design, construction and commissioning of the ATCA 12-mm system.

%\bibliographystyle{apj} 
%\bibliography{journals,psrrefs,modrefs,crossrefs}

\end{document}